\documentclass[aps,prf,preprint,amssymb,superscriptaddress]{revtex4-2}

\usepackage{graphicx}
\usepackage{natbib}
\usepackage{hyperref}
\usepackage{multirow}
\usepackage{amsmath}      % AMS Math
\newcommand{\be}{\begin{equation}}
\newcommand{\ee}{\end{equation}}
\newcommand{\bea}{\begin{eqnarray}}
\newcommand{\eea}{\end{eqnarray}}

\newcommand{\del}{\nabla}

\usepackage{color}

\begin{document}

%\title{Resolved Particulate Rayleigh-B\'{e}nard Convection}
\title{Particle Re-Suspension in Two-Phase Dispersed Rayleigh-B\'{e}nard Convection}

\author{Xianyang Chen}
\affiliation{Department of Mechanical Engineering, University of Houston, Houston TX 77204, USA}

\author{Rodolfo Ostilla Monico}
\affiliation{Dpto. Ing. Mec\'{a}nica y Dise\~{n}o Industrial, Escuela Superior de Ingenier\'{i}a, Universidad de C\'{a}diz, 
%Av. de la Universidad de C\'{a}diz 10, 
11519 Puerto Real, Espa\~{n}a }

\author{Daniel Floryan}
\affiliation{Department of Mechanical Engineering, University of Houston, Houston TX 77204, USA}

\author{Andrea Prosperetti}
\email[]{aprosper@central.uh.edu}
\affiliation{Faculty of Science and Technology,  University of Twente, 7500 AE Enschede, The Netherlands }
\affiliation{Department of Mechanical Engineering, Johns Hopkins University, Baltimore MD 21218, USA}
\affiliation{Department of Mechanical Engineering, University of Houston, Houston TX 77204, USA}

\graphicspath{{figures_JFM_reject/}}

\begin{abstract}

The process by which particles are entrained by the fluid in Rayleigh-B\'{e}nard convection is studied by means of particle-resolved numerical simulations in a periodic domain at a Rayleigh number of $10^7$. The fluid Prandtl number is 1 and the particle-to-fluid density ratio 1.1.
The results show that the horizontal velocity field near the bottom of the cell accumulates particles in heaps, or `dunes', at the base of ascending plumes. The dunes deflect the incoming flow, conferring to it a vertical velocity component which entrains the particles up the dune and into the plume. An experimental observation of this mechanism was briefly reported by Solomatov {\em et al.} ({\em Earth Planet. Sc. Lett.} {\bf 120}, 387, 1993) but has not been considered further in the literature. By its very nature, such a process cannot be simulated by the point particle model. The final particle load carried by the convection depends both on the available gravitational energy of the fluid and on the effectiveness of the re-suspension mechanism. 

\end{abstract}

\maketitle

\section{Introduction}

Rayleigh-B\'{e}nard natural convection with suspended particles is a situation encountered in a variety of settings such as air circulation in rooms and other built spaces \cite[see, e.g.,][]{Balachandaretal20,Rayeganetal23}, the ocean and the atmosphere ~\cite[see, e.g.,][]{Steinetal15,Autaetal17}, metal-production processes \cite[see, e.g.,][]{JarvisWoods94} and  magma chambers and other planetary processes \cite[see, e.g.,][]{ElkinsTanton12}. 
%Furthermore, just as single-phase Rayleigh-B\'{e}nard convection has played an important role in advancing our understanding of hydrodynamic stability, convective heat transfer and turbulence~\cite[see e.g.][]{*}, it may be expected that the study of the two-phase version of the problem may foster progress on important scientific and technological problems such as turbulent two-phase flow, heat transfer in particulate systems and the transport of finely dispersed material, to name a few areas. 

In Rayleigh-B\'{e}nard flow the fluid streamlines are, at least approximately, closed, and it might be thought that suspended particles could describe closed orbits remaining indefinitely entrained in the fluid circulation. The time-dependence and  chaotic nature of the flow, however, confers to particles heavier than the fluid a non-zero probability of settling at the bottom of the convection cell so that all negatively buoyant particles might be expected to eventually settle. Experimental evidence does not conform with this expectation since it shows that it is possible for particles to be re-suspended in the convecting layer after deposition \cite[see, e.g.,][]{MartinNokes89,Solomatovetal93}. There is also indirect evidence from the geological record as continuous particle settling would be expected to produce a strong compositional differentiation, which is not always observed. Again, this fact suggests the presence of re-suspension processes \cite{SolomatovStevenson93,ElkinsTanton12}. 

Several computational studies have examined the matter on the basis of Euler-Lagrange models with point particles~\cite[see, e.g.,][]{Parketal18,Patockaetal22}. By their very nature, these models cannot include a realistic account of re-suspension since, once particles are deposited on the bottom of the domain where the fluid velocity vanishes by the no-slip condition, there is no mechanism to re-suspend them in the convecting fluid. ~\citet{Parketal18} artificially re-injected settled particles in the fluid layer occupying the bottom 10\% of their computational domain and found that sufficiently light particles could then be re-suspended even in a convective flow with a Rayleigh number as low as $2\times 10^6$, but the degree to which their procedure mimics the actual re-suspension mechanism is uncertain. 

It appears that a necessary requirement for a computational approach to convincingly elucidate the re-suspension process is the ability to account for the particle finite size as done in the present paper. We simulate 500 
particles in a standard three-dimensional Rayleigh-B\'{e}nard convection cell with periodicity conditions in the horizontal directions and a  Rayleigh number of $10^7$. We also carry out simulations of up to 1000 particles in a quasi-two-dimensional domain, also with periodicity conditions in the horizontal directions. Our results are in %perfect 
qualitative agreement with the experimental observations reported by~\citet{Solomatovetal93}, who noticed that tangential stresses as considered in the Shields mechanism~\cite[see, e.g.,][]{Ouriemietal07} may be insufficient to entrain deposited particles. Rather, the circulating flow, which is nearly horizontal close to the cell bottom, causes the accumulation of the settled particles into structures that they refer to as  `dunes'. These structures cause the bottom fluid velocity to acquire a component in the vertical direction, which is able to drag the particles upward, placing them in ascending fluid plumes.

\section{Mathematical model and dimensionless parameters}

The fluid behavior is governed by the constant-properties Navier-Stokes equations in the Boussinesq approximation~\cite[see, e.g.,][]{Ahlersetal09}. The base of the computational domain is maintained at a temperature $T_h$ and its upper surface at a temperature $T_c<T_h$. The fluid density is assumed to depend linearly on temperature according to  $\rho(T) = \rho_0[1-\beta (T-T_0)]$, in which $T_0=\frac{1}{2}(T_h+T_c)$ and $\rho_0$ is the fluid density at $T=T_0$; $\beta$ is the constant thermal expansion coefficient. The Rayleigh $Ra$ and Prandtl $Pr$ numbers are defined by  
\be
 Ra = \frac{g\beta (T_h-T_c)H^3}{\nu_f D_f} \qquad {\rm and} \qquad Pr = \frac{\nu_f}{D_f} \,, 
 \ee
 with $g=|\mathbf{g}|$ the acceleration of gravity, $H$ the height of the cell and $\nu_f$ and $D_f$ the fluid kinematic viscosity and thermal diffusivity. 
 
Together with the cell aspect ratio, such a characterization would be sufficient for single-phase flow, but the presence of particles introduces several other important parameters. The single-particle Reynolds number $Re_p$ is defined as $Re_p=d_pU_{term}/\nu_f$, in which $d_p$ is the  particle diameter and $ U_{term}$ the particle terminal velocity. Adopting for the drag coefficient the well-known Schiller-Naumann form~\cite[see, e.g.,][]{Cliftetal78} $C_D=(24/Re_p)[1+0.15 \,Re_p^{0.687}]$, we find 
\be
U_{term} = \left(1-\frac{\rho_0}{\rho_p}\right) g \tau_p \qquad {\rm with} \qquad 
 \tau_p= \frac{\rho_pd_p^2}{18\rho_0 \nu_f (1+0.15 Re_p^{0.687})} \, ,
\ee
the particle characteristic mechanical time~\cite[see e.g.][]{Luccietal11}; $\rho_p$ is the constant particle density. 
The mechanical Stokes number $St$ (as opposed to the thermal Stokes number, introduced below)  is defined by  
\be
 St= \frac{\tau_p U_f}{H} \qquad {\rm with} \qquad  U_f= \sqrt{g\beta (T_h-T_c)H} \,,
 \ee
the fluid free-fall velocity. Pertinent values of these and other particle parameters  are listed in Table~\ref{tabprop}.
 
The thermal Stokes number $St_{th}$, necessary to characterize the particle response time to temperature changes in its neighborhood, is defined by
\be
 St_{th} = \frac{\tau_{p,th} U_f}{H} \qquad {\rm with} \qquad \tau_{p,th}= \frac{d_p^2\rho_p c_{pp}}{6k_fNu_p} \, ,
 \ee
the particle thermal response time; here $c_{pp}$ is the particle specific heat, $k_f$ the fluid thermal conductivity and $Nu_p$ the particle Nusselt number~\cite[see e.g.][]{Parketal18} estimated with the Ranz-Marshall correlation $Nu_p=2+0.6 Pr^{1/3}Re_p^{1/2}$~\cite[see, e.g.,][]{Cliftetal78}.  Note that this and the  Schiller-Naumann correlations are only used to quantify the characteristic time scales of the particles. In the simulation the hydrodynamic force and the particle-fluid heat transfer are calculated from first principles.

We assume that the thermal diffusion time in a particle, of the order of $(d_p/2)^2/D_{p,th}$, with $D_{p,th}$ the particle thermal diffusivity, is smaller than $\tau_{p,th}$, which requires that $k_p/k_f > \frac{3}{2} Nu_p$, with $k_p$ the particle thermal conductivity. The value of $Nu_p$ in the present simulations is 3.89. This inequality therefore is satisfied if the particle thermal conductivity is a bit more than an order of magnitude greater than that of the fluid. This assumption permits us to use the lumped capacitance approximation for the particles writing 
\be
 m_pc_{pp} \frac{dT_p}{dt} = k_f \oint_{s_p} \nabla T \cdot \mathbf{n}_pds_p \,,
 \label{parten}
 \ee
 in which $\mathbf{n}_p$ is the outwardly directed unit normal at the particle surface $s_p$. Appendix B describes an estimate of particle-particle heat transfer upon collision and shows that this process is negligible in our conditions even when the particle thermal conductivity is large enough to justify the lumped capacitance approximation.
 % dr Wheat eye dr
 %These are the only other papers with particle-resolved simulations in Rayleigh-B\'{e}nard convection that we have found, but the much smaller Rayleigh numbers considered by these authors and the large differences in several parameters between their study and ours make a comparison impossible.

Several other dimensionless parameters are necessary to fully characterize our simulations. To the ones already listed, we may add the following ones
\be
  \frac{\rho_p}{\rho_0} \, ,\qquad \frac{c_{pp}}{c_{pf}} \,,\qquad \frac{H}{d_p} \,, \qquad \alpha=\frac{\frac{\pi}{6} d_p^3N_p}{V}\,,
  \qquad E_* = \frac{\nu_f^2 E}{\rho_0 (1-\sigma^2) g^2 d_p^4} \, ,
  \label{defvarE}
  \ee
  Here $\sigma$ is the Poisson ratio, $c_{pf}$ the constant fluid specific heat, $\alpha$ the particle volume fraction in the cell of volume $V$ and $E$ the Young modulus of the particle material that enters the collision model described in Appendix A. The parameter space is therefore considerably enlarged by the particles which renders a detailed investigation problematic. Another consequence of this large number of parameters is that  it is next to impossible to isolate effects. For example, a change of $\rho_p/\rho_0$ would affect the particle Reynolds number $Re_p$ and both Stokes numbers. A change to the mechanical Stokes number by changing $H$ would cascade through several dimensionless groups.  Adopting units such that $H=1$ and $T_h-T_c=1$, a  common procedure in single-phase studies, would require to make specific choices for $g$ and $\nu_f$ which would have major effects on $Re_p$ and other parameters. These considerations have placed strong constraints on our choice  of parameters that has been dictated by the selection of a small region of parameter space where we could expect some interesting results more than by the selection of a realizable physical system that could be  investigated experimentally. 
  
A related problem which is common in single-phase Rayleigh-B\'{e}nard convection but is seldom explicitly discussed in the literature is that the choice of parameters for which simulations are possible very often does not correspond to experimentally  realizable situations. For example, for water at 20 $^\circ$C with $T_h-T_c=10$ K, $Ra=10^7$, as in the present work, would correspond to a cell with a height of only 41 mm. For air the cell height becomes 213 mm, but it is difficult to imagine solid particles with a density only 10\% greater than air and, for this Rayleigh number, the convection is not strong enough to suspend much heavier particles. The `obvious' solution would be to increase the Rayleigh number, but the Reynolds number of convection, on which particle suspension ultimately depends, is proportional to $\sqrt{Ra}$ so that an increase by an order of magnitude would change things very little, while an increase by two orders of magnitude by increasing $H$ would increase the total number of computational nodes by the same factor and place the resolved simulation of  suspended particles completely outside the feasibility range. Increasing the Rayleigh number by increasing $g$ or decreasing $\nu_f$ would increase the particle Reynolds number making them too heavy for suspension. These examples can easily be multiplied. The objective of this paper is to present one of the first resolved-particle simulations to begin the exploration of what happens when one leaves behind the point-particle models on which all the existing work on particulate Rayleigh-B\'{e}nard convection (save that of Kajishima and collaborators, see e.g.~\cite{Guetal19,Takeuchietal19}, who simulated $Ra=10^5$) has been invariably based until now.

Two cells were used in the simulations, in both cases with periodicity and no-slip boundary conditions applied on the vertical and horizontal faces, respectively. One cell was cubic, with aspect ratio 1 and a side $H=20\, d_p$. In this cell we ran simulations for about $180 \, H/U_f$ time units, preceded by $20\,H/U_f$ time units as the flow was established. The computation lasted over two months on 8 Nvidia V-100 GPUs. The second cell was quasi-two dimensional, with height $H=20 \,d_p$ and horizontal dimensions of $40\,d_p\times 3 \,d_p$. Use of this latter cell was dictated by a desire to further study the re-suspension mechanism discovered in this work and to permit a (limited) exploration of the effect of particle number in a reasonable time.

We choose $d_p/2.5$ for the length scale and $(d_p/2.5)^2/\nu_f$ for the time scale. In these units the dimensionless free-fall velocity  is $d_pU_f/(2.5 \nu_f) = \sqrt{4000}\simeq 63.25$. The temperature at the cell base is fixed at $T_h/(T_h-T_c)=+0.5$ and at the cell top at $T_c/(T_h-T_c)=-0.5$; furthermore $\beta(T_h-T_c) =0.4$ and $(d_p/2.5)^3 g/\nu_f^2 =200$.

\begin{table}
\centering
\begin{tabular}{| c |  c| c|c| r| c| c|c|c|}
\hline
$\rho_p/\rho_0$ & $Re_p$ & $Nu_p$ & $St$ & $St_{th}$ & $U_{term}/U_f$ & $\rho_p/\rho(T_c)$ & $\rho_p/\rho(T_h)$ & $c_{pp}/c_{pf}$\\
\hline 
1.1 & 10.0 & 3.89 & 0.28 & 0.37 & 0.063 & 0.917&1.38 & 1\\
%1.2 & 16.9 & 0.26 & 0.35 & 0.11 &1.00 & 1.50 \\
%1.4 & 28.0 & 0.25 & 0.36 & 0.18 & 1.17 & 1.75\\ 
\hline
\end{tabular}
\caption{The value of some particle parameters  used in the present simulations. The fluid parameters were held fixed: $Ra=10^7$, $Pr=1$,   $\beta(T_h-T_c) =0.4$.}
\label{tabprop}
\end{table}

\section{Numerical method}

The mathematical problem for the fluid is solved numerically by the {\sc physalis} method, which has been thoroughly described in several earlier publications~\cite[see, e.g.,][]{SierakowskiProsperetti16,Wangetal17}.  The method relies on the use of analytic general solutions for momentum and energy to replace boundary conditions on the particle surfaces by equivalent conditions on the neighboring nodes of a regular Cartesian grid. 
This procedure is made possible by the no-slip condition which, in the particle rest frame, makes the fluid velocity very small in the immediate neighborhood of the particle surface so that convection can be neglected. In the same very limited region (between the particle surface and the closest grid nodes) the time derivative of both velocity and temperature is also  neglected so that the energy equation, for example, is approximated as $\del^2T=0$ and the momentum equations are approximated by the Stokes equations. These approximations, which are fully justified in the references cited, make analytical solutions available for the momentum and energy equations in the form of (suitably truncated) series of vector and scalar spherical harmonics multiplied by known functions of the distance from the particle center. For example, the solution of the energy equation in the immediate neighborhood of the particle surface is approximated as
\be
 T({\bf x},t) = T_p(t) +\sum_{\ell=0}^\infty \left[\left(\frac{r}{r_p}\right)^\ell-\left(\frac{r_p}{r}\right)^{\ell+1}\right] \sum_{m=-\ell}^\ell T_{\ell m}(t)Y_\ell^m(\theta,\phi) \,,
 \label{Tex}
 \ee
in which $Y_\ell^m$ are spherical harmonics, $(r=|{\bf x}|,\, \theta,\, \phi)$ is a system of spherical coordinates centered at the particle center, $r_p$ is the particle radius and the coefficients $T_{\ell m}(t)$ are chosen so as to match the solution of the complete unsteady convection-diffusion equation at the neighboring nodes. Similar expressions are available for the general solution of the Stokes equations~\cite[see, e.g.,][]{KimKarrila91}.

Since the coefficients multiplying the terms of the expansion (\ref{Tex}) and the similar ones for velocity and pressure  are unknown, the calculation proceeds by iteration. Starting with guessed values, a suitable truncation of the series is used to assign temperature, velocity and pressure at the grid nodes closest to the particle surface. These values are used as boundary conditions to solve the (complete) Navier-Stokes and convection-diffusion equation throughout the fluid domain. New values of the coefficients are then obtained from these solutions by taking scalar products over surfaces concentric with the particles having a radius 15-20\% greater than the particle radius. The difference between the coefficients obtained in this way and the previous values is used to drive the iterations. Convergence is ultimately achieved because, for a given set of boundary conditions, the field values on the scalar-product surface away from the particle surface are compatible with the boundary conditions only if the field equations are satisfied in the intervening layer. A very useful aspect of this approach is that the coefficients of the expansions embody important physical information such as the hydrodynamic force, hydrodynamic torque and heat transfer rate for each particle, thus avoiding the need to integrate the numerical result for the fluid stress and heat flux over the particle surface. For example, it is readily shown upon substituting the expansion (\ref{Tex}) in (\ref{parten}) that
\be
 k_f \oint_{s_p} \nabla T \cdot \mathbf{n}_pds_p = 2\sqrt{\pi} r_p k_f T_{00}(t) \, .
 \label{Qpart}
 \ee

Given the limited Reynolds number of the fluid-particle motion that we encounter in the present paper (see Table~\ref{tabprop}), on the basis of grid convergence studies carried out in earlier works~\cite[][]{SierakowskiProsperetti16,Wangetal17,Willenetal17,Willenetal19}, for the present simulations we have used uniform Cartesian grids with 8 nodes per particle radius. Again on the basis of past experience,  
the summations in the general solutions for the energy and momentum equations have been truncated to the first 16 and 46 terms, respectively.  
%In the case of heavier particles with $\rho_p/\rho_0=1.2$ and 1.4, their greater accumulation on the cell bottom forced us to truncate the summations to 2 and 5 terms for energy and momentum. In these cases the simulations are somewhat under-resolved but the results are indicative of what may be expected with a more refined numerical method. 

Equation (\ref{contt}) of Appendix B gives the time $t_c$ during which two colliding particles remain in contact. After non-dimensionalization this expression becomes
\be
 \frac{\nu_ft_c}{d_p^2} = 1.69  \left(\frac{5\pi}{1296}\right)^{2/5}  \frac{1 }{Re_{St}} 
 \left[\frac{\rho_p}{\rho_f} \left(\frac{\rho_p}{\rho_f}-1\right)^2  \frac{1}{E_*} \right]^{2/5} \, ,
 \label{collt}
 \ee
in which $Re_{St} = \frac{1}{18}(\rho_p/\rho_f-1)gd_p^3/\nu_f^2$ is the particle Reynolds number based on the Stokes velocity, (rather than the terminal velocity, which includes a Reynolds-number dependent correction that must be calculated numerically)  and the dimensionless Young modulus $E_*$ is defined in (\ref{defvarE}).  In our simulations we use $E_*=10^6$ which, for example for water, with  $\sigma \simeq \frac{1}{2}$ and $d_ p$ between 1 and 0.1 mm, would correspond to $E$ between $10^5$ and $10^9$ Pa. With this value the mechanical characteristic time $\tau_p$ is about 12 times the contact time and the time step used in the calculations about 23 times smaller. Collisions can therefore be resolved in the simulations.

\section{The re-suspension mechanism}

\begin{figure}
\centering
\includegraphics[width=0.9\textwidth]{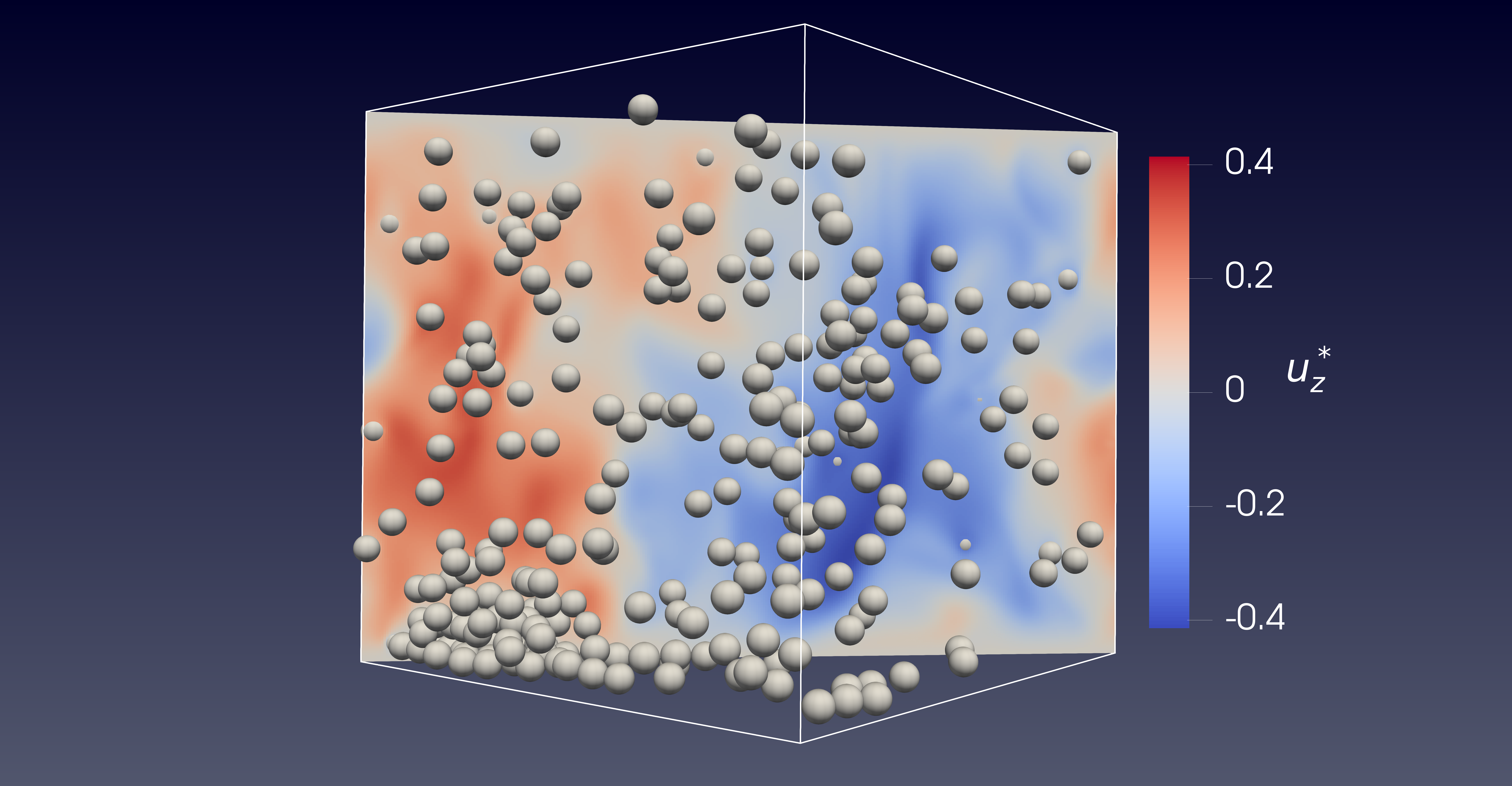}
\caption{Snapshot of a simulation with 500 particles with a density $\rho_p/\rho_0 = 1.1$ and an average volume fraction of 3.27\%. The Rayleigh number is $10^7$ and the Prandtl number is 1. The color on the plane cutting diagonally through the cell indicates the normalized vertical velocity $u_z^*=u_z/U_f$.}
\label{f500}
\end{figure}

At a Rayleigh number comparable to the one of our simulations, $Ra=10^7$, the flow structure in single-phase Rayleigh-B\'{e}nard convection consists of buoyant plumes rising from the warm boundary layer at the bottom of the cell and cold plumes descending from the cold boundary layer forming along the cold top. These plumes drive a large-scale circulation, also known as mean wind, with a size of the order of the plate separation and a nearly vertical orientation~\cite[see e.g.][]{BrownAhlers06}. In a doubly periodic domain, the plumes wiggle around but do not move very much.  At our Rayleigh number the flow is mildly turbulent with a Reynolds number $HU_f/\nu_f \simeq 3200$. In the range $3\times 10^8\lesssim Ra \lesssim 10^{11}$ the effect of the Prandtl number for fixed $Ra$ is less than 2\% for $4 \lesssim Pr \lesssim 34$. 
Our simulations add $N_p=500$ particles, achieving a mean volume fraction of 3.27\%, but the character of the flow remains largely unchanged.

Figure~\ref{f500} is a snapshot from a simulation in which $N_p=500$ particles with %a density $\rho_p/\rho_0=1.1$ and 
a mean volume fraction of 3.27\%  are suspended in a Rayleigh-B\'{e}nard flow with $Ra=10^7$ and $Pr=1$.  Movie 1, showing the time development of this flow, is available at {\tt https://doi.org/***}. 
%Periodicity and no-slip boundary conditions are applied on the vertical and horizontal faces of the cell, respectively. 
%The cell has aspect ratio 1 and the length of each side is 20 particle diameters so that, with $d_p=2.5$ in dimensionless units, the range of all coordinates is $[-25,25]$. The time scale is $d_p^2/\nu_f$ and the velocity scale is $2.5\, \nu_f/d_p$, in terms of which $U_f = \sqrt{4000}\simeq 63.25$. The temperature at the cell base is fixed at $T_h=+2$ and at the cell top at $T_c=-2$, $\beta =0.1$ and $g=200$.  The ratio of the particle diameter to the common viscous and thermal boundary layer thicknesses  is $d_p/\sqrt{\nu_f H/U_f} \simeq 2.8$. 
In the figure the color on the vertical plane cutting diagonally through the cell is the fluid vertical velocity and shows ascending and  descending plumes, in red and blue, respectively. In addition to the suspended particles, a point to note, which is the focus of this article, are the particles on the bottom of the cell and, in particular, the particles heaped up in a `dune' in a corner of the domain.  

\begin{figure}
\centering
\includegraphics[width=0.7\textwidth]{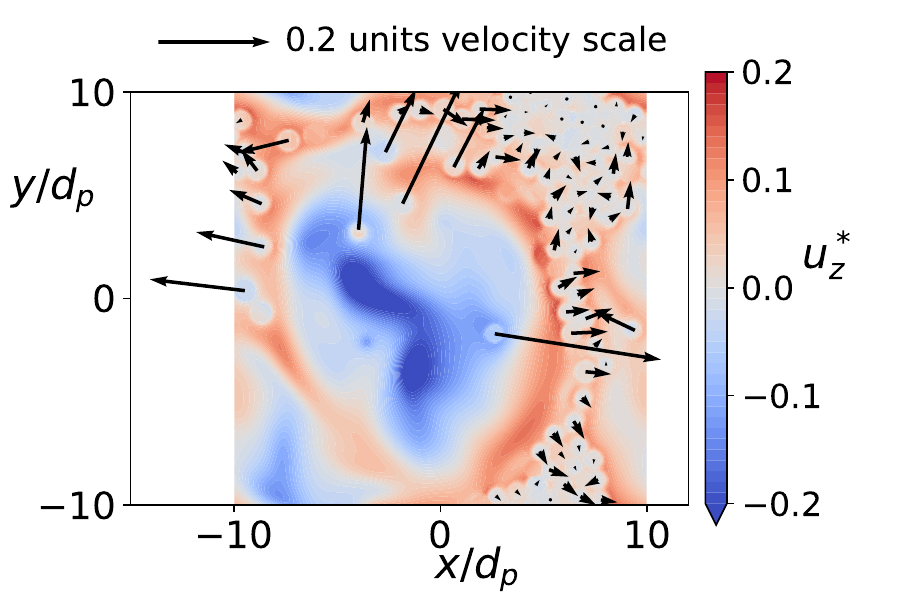}
\caption{The color is the normalized fluid vertical velocity  $u_z^*=u_z/U_f$ on a horizontal plane one diameter above the cell base. The rms absolute value of the vertical velocity at this distance from the cell bottom is 2.6 times smaller than the horizontal velocity; one radius above the bottom it is 5.5 times smaller. Arrows indicate the horizontal velocity of  particles with a center within one diameter of the base.}
\label{fqui}
\end{figure}

The mechanism by which a dune is formed is illustrated in figure~\ref{fqui}, which is a snapshot of the vertical velocity on a horizontal plane one diameter above the base of the cell. Blue and red colors correspond to negative (descending) and ascending fluid velocities, respectively.  In the figure, arrows are attached to particles with a center within one particle diameter of the bottom and their length indicates the horizontal velocity of each particle. Near the bottom of the cell the fluid flow is approximately horizontal, with the rms of the vertical velocity on the plane of the figure about 2.6 times smaller than the horizontal velocity, and 5.6 times smaller on a horizontal plane located one particle radius below it. By continuity, this horizontal flow is directed from regions under the descending plume to regions under the ascending one which, at this time instant, forms a roughly cylindrical curtain encircling the descending plume. Particles that have settled on the bottom are exposed to this flow, which pushes them toward the base of the ascending plume where they accumulate. The drag imposed by the flow is sufficiently strong that particles form heaps as in figure~\ref{f500}.

%\begin{figure}
%\centering
%\includegraphics[width=0.45\textwidth]{Figures/zmean_500.pdf} 
%\includegraphics[width=0.45\textwidth]{Figures/sediment_pdf2.pdf}
%\caption{zmean\_500.pdf}
%\label{fzmn}
%\end{figure}

\begin{figure}
\centering
\includegraphics[width=0.47\textwidth]{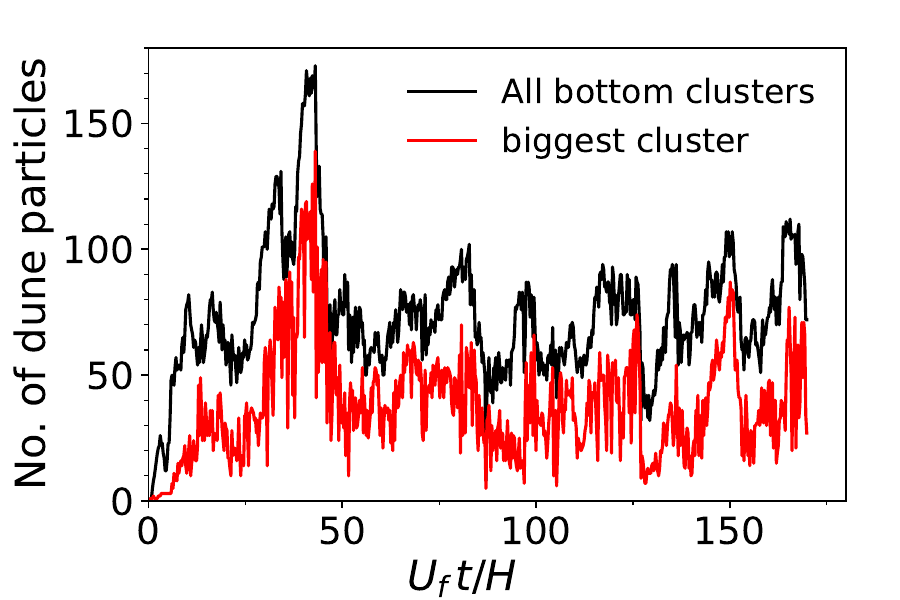}
\includegraphics[width=0.47\textwidth]{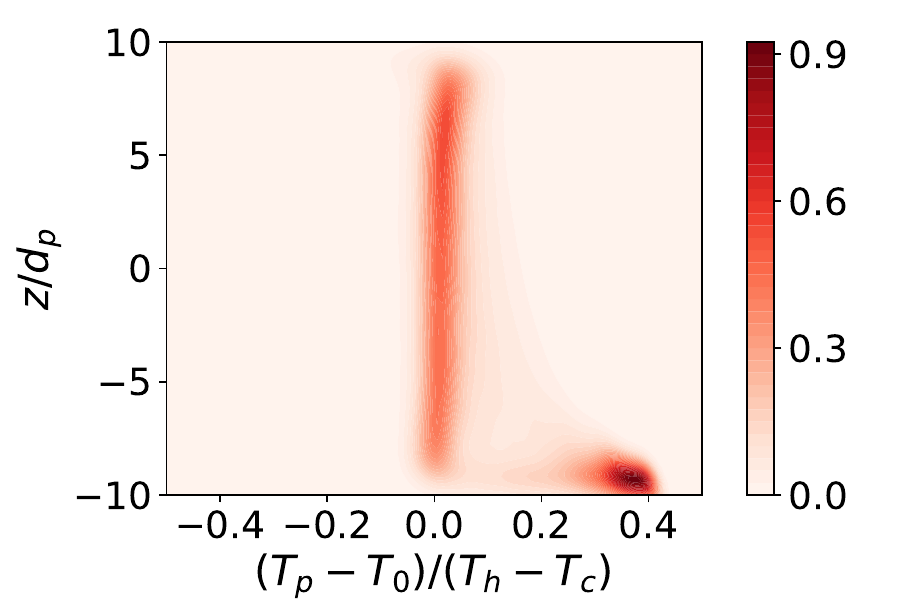}
\caption{$(a)$ The red line shows the number of particles belonging to the largest bottom cluster (dune) vs. time. The black line is the total number of particles, clustered, or unclustered, on the cell bottom. Both lines refer to the simulation of figures~\ref{f500} and~\ref{fqui}. $(b)$ Joint probability density function of the particle vertical position and normalized temperature.}
\label{fsed500}
\end{figure}

To quantify the number of particles in dunes at any given time, we use a variant of the union-find algorithm based on the distance between the centers of particle pairs. A particle is declared to belong to a cluster when the distance of its surface from that of another particle belonging to the cluster is less than or equal to zero.  Particles that are classified as bottom particles either belong to clusters (of whatever size) resting on the bottom, or are isolated particles touching the bottom.
%Voronoi tessellation of the set of points constituting the particle centers. For each particle, we calculate the total volume fraction that it occupies in its Voronoi cell. To decide whether a particle belongs to a dune we threshold this volume fraction at 35\%. This procedure correctly identifies particles that are surrounded essentially on all sides by other particles but it may not include particles on the surface layer of a dune, and can therefore somewhat under-estimate the particles forming the dune. 
The black line in figure~\ref{fsed500}$(a)$ shows, as a function of time, the total number of bottom particles while the red line is the number of particles in the largest dune. The simulations are started by keeping the particles fixed in random positions until the convection is established, at which point, taken as $t=0$, they are released. The accumulation shows a temporary peak, still part of the initial transient, and eventually stabilizes fluctuating around a value of about 71 bottom with a standard deviation of 16. The average number of particles in the largest dune is $37 \pm 16$.  
%as the particles, subjected to the circulating flow, accumulate in the dune are suspended, fall back to the domain bottom and 
 
%A relation between the temperature of the bottom particles and their mutual distance can be established by carrying out a two-dimensional Voronoi tessellation of the set of particle centers within one diameter from the cell bottom. 
Figure~\ref{fsed500}$(b)$ shows the joint probability density function of the normalized particle temperature and the normalized vertical position. The figure shows that the warmest particles are concentrated near the bottom of the cell. Most of these particles belong to the largest dune and help reinforce the ascending plume. Particles distributed throughout the height of the cell mostly have a temperature close to the average value $\frac{1}{2}(T_h+T_c)$. Near the bottom we see slightly cooler particles that have fallen from the colder region while the particles near the top are slightly warmer having been transported by the rising plume. The situation has some analogies with that described in~\citet{Morizeetal17}, who studied particle re-suspension by thermal plumes in a Rayleigh-B\'{e}nard-type setup. Their cell, filled with a water-calcium chloride mixture,  initially had a uniform particle layer at its bottom. At the beginning of the experiment, heating began to be applied over the central part of the cell bottom.  When the liquid acquired sufficient buoyancy, a plume formed and was able to carry the particles upward.

%The pdf of the particle residence time in a layer adjacent to the bottom of the cell is shown in figure~\ref{*}. Particles are deemed to belong to the layer if their center is within two diameters from the bottom.  The most bottom probable residence time is * time units and the mean residence time is calculated to be *. There is a very small number of particles that spends more than * time units on the bottom of the domain, which implies that nearly all particles get picked up and become suspended in the fluid at one time or another. Thus, over the total duration of the simulation, about * time units, each particle gets resuspended * times on average. It is to the nature of this process that we now turn. 

%two peaks, one near zero and one near 100\%. The latter particles are those that end up in the interior of a dune, which have a very small chance of escaping and be resuspended. The peak on the left of the figure shows that many particles spend a relatively small amount of time on the bottom of the domain, which implies the presence of an active process able to re-suspend them. 

\begin{figure}
\centering
\includegraphics[width=0.65\textwidth]{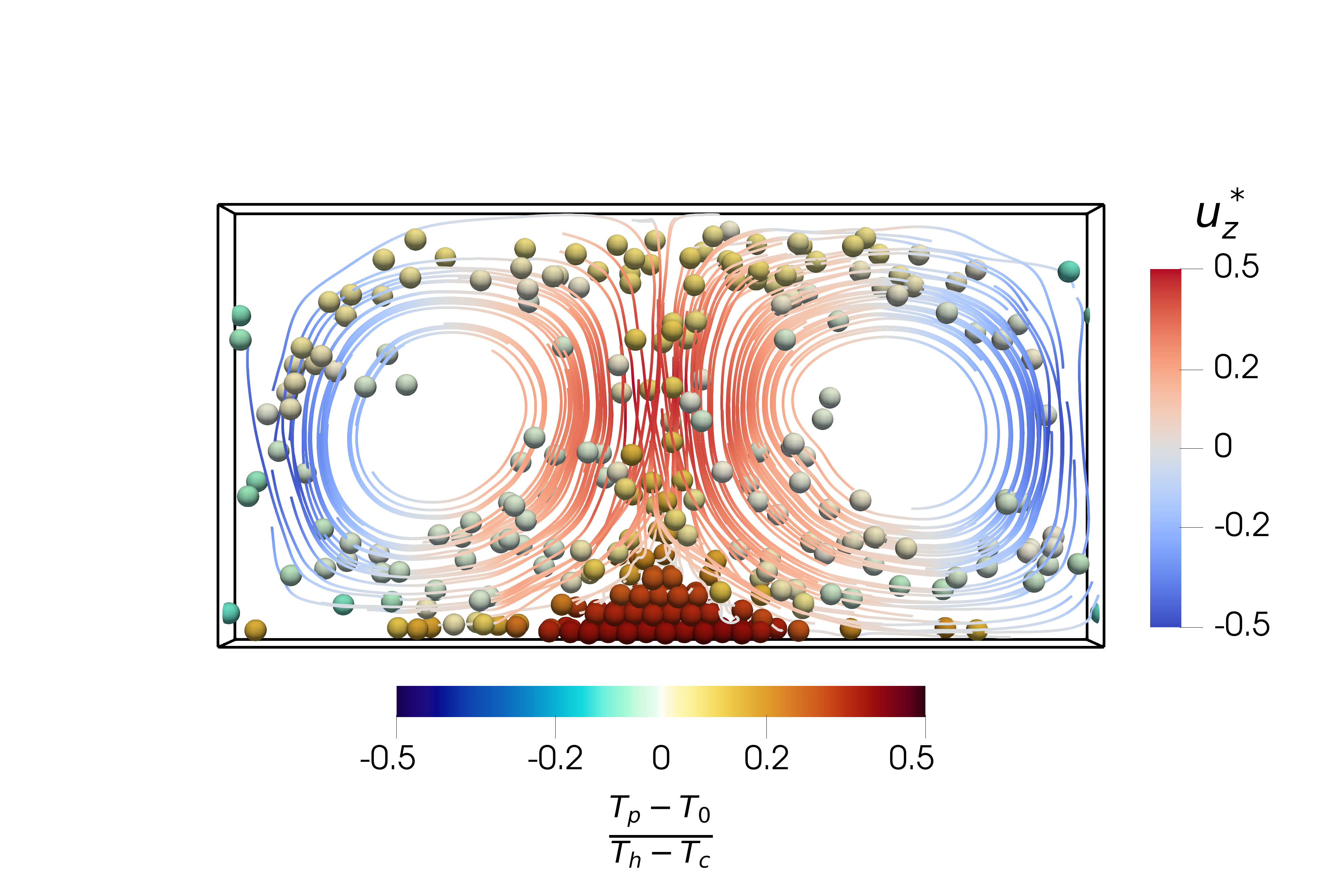}
\caption{Instantaneous streamlines and particle dune for the flow in a cell with aspect ratio 2, $Ra=10^7$ and $Pr=1$; the total number of particles is 300 and $\rho_p/\rho_0=1.1$. The dune has been placed at the center of the domain bottom by taking advantage of the lateral periodicity of the simulations. 
The streamlines are colored by the normalized vertical velocity  $u_z^*=u_z/U_f$. Movie 2, available at {\tt https://doi.org/***},  shows the time history of this simulation.} 
\label{fstrea}
\end{figure}

\begin{figure}
\centering
\includegraphics[width=0.48\textwidth]{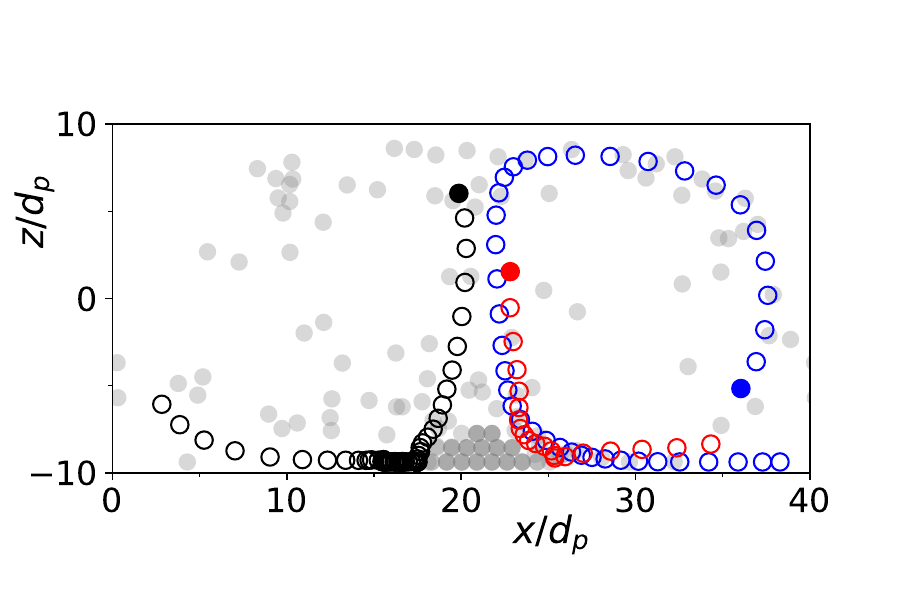}
\includegraphics[width=0.48\textwidth]{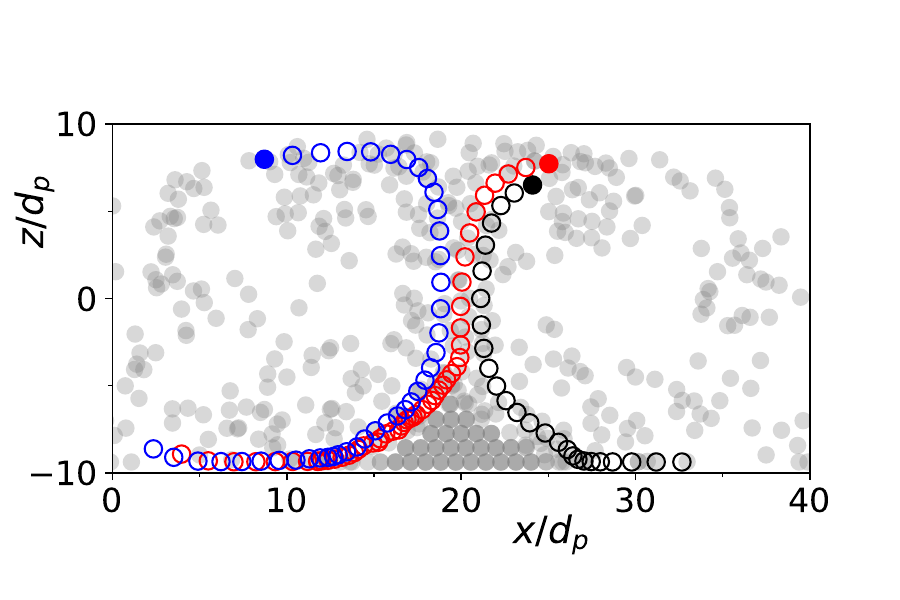}
\caption{$(a)$  
Illustration of the re-suspension mechanism for $N_p=150$ (left) and $N_p=500$. The hollow circles show the positions of the black, blue and red particles at earlier times. The particle dune imparts a vertical component to the fluid velocity which entrains particles into the ascending stream above the dune.
%$(b)$ Number of particles vs. normalized time in each layer of the dune formed in the case $\rho_p/\rho_0=1.2$ with $N_p=150$. 
%%$(b)$ Number of particles $N_{p,d}$ forming a dune divided by the total particle number $N_p$ vs. normalized time in the quasi-two-dimensional simulations. 
%In ascending order, the lines correspond to $N_p=150,\, 300, \,500$ and 1000.  
} 
\label{fsed150}
\end{figure}

The physics of the re-suspension process becomes clearer 
%(and the computational time shorter) 
if we consider a quasi-two-dimensional cell with aspect ratio 2. The height is 20 $d_p$ as before, but the dimensions of the base are 40 $d_p \times 3 \, d_p$. Doubling the length of the base has the effect of producing two counter-rotating convection cells, as shown in figure~\ref{fstrea}.  The two cells form a dune by the same mechanism described in connection with figure~\ref{fqui}. 
The mechanism by which the dune favors particle re-suspension is illustrated in figure~\ref{fsed150} for $N_p=150$ (left) and $N_p=500$: the hollow circles show the positions of the black, blue and red particles at earlier times and trace out their trajectory as the particles fall to the cell bottom, are pushed toward the dune, dragged up its slopes and ultimately re-suspended into the ascending stream. Unlike the previous case, in this quasi-two-dimensional cell essentially all the bottom particles belong to a  single large dune.

\begin{figure}
\centering
\includegraphics[width=0.48\textwidth]{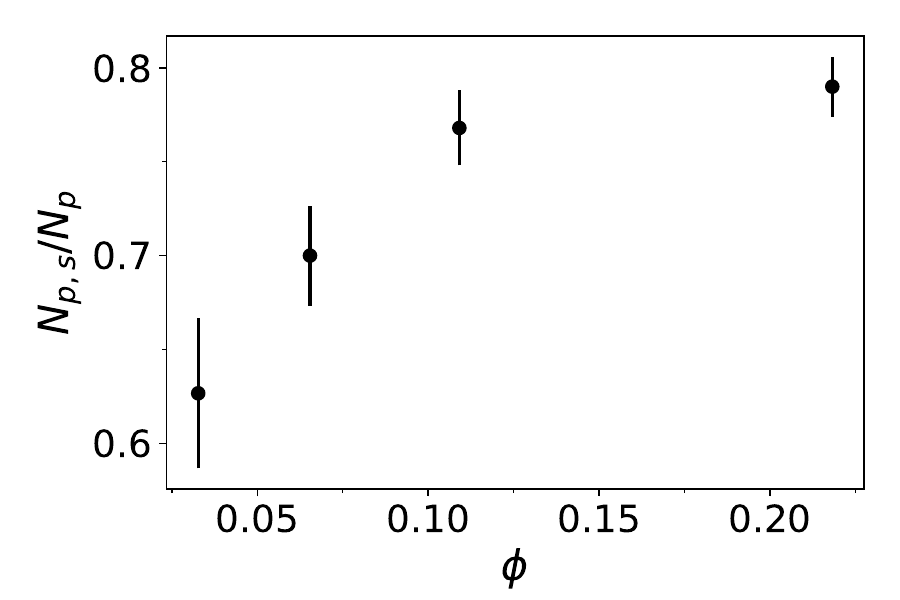}
\includegraphics[width=0.48\textwidth]{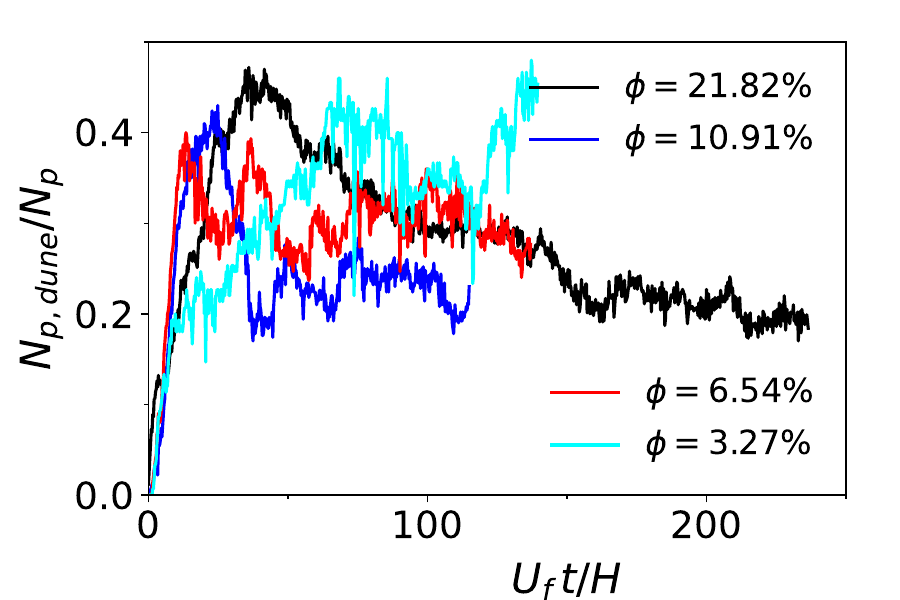}
%\includegraphics[width=0.47\textwidth]{Figures/Nps_Np_rhop1.1.pdf}
%\caption{Illustration of the resuspension mechanism for $N_p=150$ (left) and $N_p=300$, both with $\rho_p/\rho_0=1.1$. The hollow circles show the positions of the black, blue and red particles at earlier times. The particle dune imparts a vertical component to the fluid velocity which entrains particles into the ascending stream above the dune. } 
\caption{$(a)$ Fraction of suspended particles $N_{p,s}/N_p$ vs. the average particle volume fraction. $(b)$ Number of particles $N_{p,d}$ forming a dune divided by the total number of particles $N_p$ vs. normalized time in the quasi-two-dimensional simulations. } 
\label{fNpsNp}
\end{figure}

The average number of suspended particles, $N_{p,s}$, for the simulations with $N_p=150,\, 300,\, 500$ and 1000 is given in Table~\ref{tcomp1} and, normalized by $N_p$, is shown in figure~\ref{fNpsNp}$(a)$ vs. the average particle volume fraction $\phi$; the vertical bars are the standard deviations. 
% and the dashed line -- a simple cubic fit  through the data --  serves as a guide to the eye. 
The rapid increase of $N_{p,s}/N_p$ for the first three data points indicates that the number of suspended particles increases more than proportionally to the total number of particles. It must be concluded that the number of suspended particles is not only limited by the load-carrying capacity of the recirculating stream, but also by the rate at which particles can be re-suspended, with the efficacy of the re-suspension mechanism increasing with $N_p$. A possible explanation is the increase of the dune size with particle number shown in figure~\ref{fsed150}. 
For $N_p=1000$, we find that the dune is largest (not shown, but see Movie 3 at {\tt https://doi.org/***}), but the fraction of suspended particles is approximately the same as for $N_p=500$ (figure~\ref{fNpsNp}$a$). The flattening of the data for $N_{p,s}/N_p$ vs $\phi$ suggests a saturation of the load-carrying ability of the convection. 
The number of particles forming the dune, which is the counterpart of the number $N_{p,s}$ of suspended particles, is shown as a function of time in figure~\ref{fNpsNp}$(b)$ for the simulations with 150, 300, 500 and 1000 particles. The system behavior for $N_p=150,\,300$ and 500 is very similar. For $N_p=1000$ the particles initially slump to the cell bottom forming a continuous layer which, little by little is eroded forming a dune over a relatively long period of time.

%\begin{figure}
%\centering
%\includegraphics[width=0.45\textwidth]{Figures/trajectory_heap_r1.2.pdf}
%\includegraphics[width=0.45\textwidth]{Figures/trajectory_heap_r1.1_300p.pdf}
%\caption{rajectory\_heap\_r1.2.pdf trajectory\_heap\_r1.1\_300p.pdf}
%\label{ftraj2}
%\end{figure}
\begin{table}
    \centering
    \begin{tabular}{|c|p{2cm}||p{2cm}|p{2cm}|p{2cm}|p{2cm}|}
        \hline
     &$N_p=500$ & $N_p=150$ & $N_p=300$ & $N_p=500$ & $ N_p=1000$ \\
          \hline
     $N_{p,s}$ & $429\pm16$ & $94\pm 6$ & $210 \pm 8$  & $384\pm10$ & $790\pm 16$ \\
       $\phi_s$ & $(2.81\pm 0.10)\%$ & $(2.05\pm 0.13)\%$ & $(4.58 \pm 0.17)\%$ & $(8.38\pm 0.22)\%$ & $(17.2\pm0.35)\%$ \\
     $E_p$ & $ 0.857\times10^5$ & $0.145\times10^5$ & $0.485\times10^5$ & $0.778\times10^5$ & $1.25\times10^5$  \\
     $E_f$ & $3.17\times10^6$ & $9.00\times10^5$  & $8.68\times10^5$ & $8.28\times10^5$ &  $7.01\times10^5$  \\
     $\Phi$ & $3.07 \times 10^6$ & $8.68 \times 10^5$ & $7.78\times 10^5$ & $7.06\times10^5$ & $5.37\times10^5$ \\
     %$W_{hd}$ & $5.95 \times 10^4$ & $2.17 \times 10^4$ & $6.58\times 10^4$ &  & $1.07\times10^5$ \\
          $\dot{W}_{hd}$ & $1.48\times10^5$  & $0.371\times10^5$ & $1.15\times10^5$  & $1.62\times10^5$  & $2.31\times10^5$ \\
 %         $W_{col}$ & $-6.13\times10^4$  & $-2.25\times10^4$  & $-6.56\times10^4$  & $-8.32\times10^4$ & $-10.63\times10^4$ \\
         $\epsilon=E_p/E_f$ & 2.76\% & 1.61\% & 5.59\% & 9.40\% & 17.8\% \\
     \hline 
     $Nu$ & 17.39 & 16.5  & 16.43  & 16.33   & 15.6 \\
     \hline
     $E_p^{Sol}$ & $3.09 \times10^5$ & $0.676\times10^5$  & $1.50\times10^5$  & $2.74\times10^5$ & $5.58\times10^5$ \\
    $\epsilon^{Sol}$ & 9.75\%  & 7.51\%  & 17.3\%  & 33.1\% & 79.6\% %&  233.3
    \\
     \hline
    $\tau_T/(\rho_p-\rho_0) gd_p$ & 0.099  &  0.096 & 0.091 & 0.086 &  0.076  \\
\hline
  %   \hline
    \end{tabular}
%    \caption{$\tau_T^2=\frac{\mu_f \Phi}{2V}$ }
      \caption{Time-averaged computed results for various flow quantities defined in section~\ref{sener}. The computed single-phase Nusselt numbers are 16.9 and 16.4  for the aspect-ratio 1 and 2 cells, respectively; $\tau_T=\sqrt{\mu_f\Phi/V}$  is the buoyancy stress scale defined in~\citet{Solomatovetal93}. }
    \label{tcomp1}
\end{table}

\section{Energetics of particle suspension}
\label{sener}

Table~\ref{tcomp1} shows the present computed results for several quantities of interest. The first two lines are the computed values for the time-averaged suspended particle number $N_{p,s}$ and corresponding particle volume fraction $\phi_s$. 

In their study, ~\citet{SolomatovStevenson93} drew attention to $E_f$, the gravitational potential energy gained by the fluid per unit time, and the corresponding potential energy for the particles, $E_p$. We define these two quantities by 
\be
 E_ p = -v_p {\bf g}\cdot \sum_{\alpha=1}^{N_p} \overline{[\rho_p - \rho(T^\alpha_p)]\mathbf{U}_\alpha} 
 %\simeq \phi V \Delta \rho gU_{term}  
 \,, \qquad  E_f =- \beta \rho_0  \mathbf{g}\cdot \overline{\int_{V_f} (T-T_0)\mathbf{u} dV_f} \, ,
 \label{Epgood}
 \ee
in which the overline denotes the time average, $v_p=\frac{\pi}{6}d_p^3$ is the common particle volume, $\bf u$ is the fluid velocity and $T_p^\alpha$ and $\mathbf{U}_\alpha$ the instantaneous temperature and velocity of the $\alpha$-th particle. A consideration of the time-averaged fluid kinetic energy balance  leads to the relation 
\be 
\Phi \equiv \frac{1}{2\mu_f} \overline{ \int_{V_f} \boldsymbol{\tau} : \boldsymbol{\tau} dV_f }
 = E_f +E_p -\dot{W}_{hd} \qquad {\rm with} \qquad \dot{W}_{hd}= \sum_{\alpha=1}^{N_p} \left(
\overline{ \mathbf{U}_\alpha \cdot \mathbf{f}_\alpha^{hd} }
+\overline{ \mathbf{\Omega}_\alpha \cdot \mathbf{l}_\alpha^{hd} } \right) \, ,
\label{EcEfWhd}
 \ee
 with $\boldsymbol{\tau}$ the viscous stress tensor and $\dot{W}_{hd}$ the work per unit time performed by the hydrodynamic force $\mathbf{f}^{hd}_\alpha$ (without the buoyancy contribution) and couple $\mathbf{l}^{hd}_\alpha$ on the $\alpha$-th particle translating with velocity $\mathbf{U}_\alpha$ and rotating with angular velocity $\mathbf{\Omega}_\alpha$. In single-phase convection $E_p=\dot{W}_{hd}=0$ and one recovers the standard relation $E_f=\Phi$ in which the integrals are over the entire cell volume $V$ rather than over $V_f$, the cell volume excluding the particles,  as here. The quantities $E_p,\,E_f,\,\Phi$ and $\dot{W}_{hd}$, as well as $E_p^{sol}$ defined in (\ref{EpSol}),  are all scaled by $\rho_0 \nu_f^3/(d_p/2.5)$.

~\citet{SolomatovStevenson93} estimated $E_f$ from the single-phase relation $E_f=\Phi$, an approximation that, in the light of the data in Table~\ref{tcomp1}, is not unreasonable. On the other hand, their estimate for $E_p$ is
\be
   E_p^{Sol} \simeq \frac{\rho_p-\rho_0}{\rho_{av}} g M_{p,s} U_{term} = \frac{\rho_p}{\rho_{av}}   \phi_s V (\rho_p-\rho_0) gU_{term}  \,,
   \label{EpSol}
   \ee
 with $M_{p,s}= \frac{\pi}{6}d_p^3 N_{p,s} \rho_p$ the mass of suspended particles and $\rho_{av} =(1-\phi)\rho_0+\phi \rho_p$ the average mixture density. Aside from the factor $\rho_p/\rho_{av}$ which, in our simulations, is not very different from 1, this relation is based on approximating the vertical component of the particle velocity $\mathbf{U}$ in (\ref{Epgood}) by the terminal velocity. While particles may be at terminal velocity relative to the fluid parcel which surrounds  them, it would seem that $\mathbf{U}$ should include the velocity of the latter as well.  The same point is made in~\citet{Sturtzetal21}. 
The omitted term is not small since, from the fact that the average vertical particle velocity must vanish, it can be deduced that its average must be equal and opposite the terminal velocity.   A comparison of $E_p$ from (\ref{Epgood})  with $E_p^{Sol}$ from (\ref{EpSol}) (with $U_{term}$ from Table~\ref{tabprop}), both calculated from the results of our simulations using for $\phi_s$ the volume fraction of the suspended particles only, is shown in Table~\ref{tcomp1}. As can be seen, the difference is in fact substantial, with $E_p^{Sol}$ consistently greater than our $E_p$. 

This difference acquires further importance if we turn to the ratio $\epsilon=E_p/E_f$ which~\citet{SolomatovStevenson93} introduced to quantify the efficiency with which the particles partake of the  gravitational energy acquired by the fluid per unit time.  Their estimate for this efficiency parameter gives it a small value, 0.6 - 0.9\%. Table~\ref{tcomp1} shows $\epsilon$ as computed with our definitions of $E_f$ and $E_p$ and $\epsilon^{Sol}$ computed from $\epsilon^{Sol} =E_p^{Sol}/\Phi$. 
The first thing to notice is the substantially larger value of both our $\epsilon$ and $\epsilon^{Sol}$ in comparison with the estimate in Solomatov \& Stevenson's paper. Secondly, $\epsilon^{Sol}$ is consistently greater than our computed value even approaching 80\% for $N_p=1000$. This value is so large as to reinforce the doubts expressed before on the way in which $E_p^{Sol}$ is calculated. 

In the paper in which they introduced the idea of dunes and their importance for re-suspension, ~\citet{Solomatovetal93} also concluded that re-suspension takes place when the parameter $\sqrt{\mu_f\Phi/V}/[( \rho_p-\rho_0) g d_p]$ is around 0.1. Our results for this quantity, also shown in Table~\ref{tcomp1}, are in good agreement with this estimate.  The particles have a very modest effect on the cell Nusselt number, which is reduced by only about 5\% by a particle volume fraction as high as 21.8\%. 

~\citet{LavorelLeBars09} used the same estimate as $E_p^{Sol}$ for the particle gravitational energy omitting the factor $\rho_p/\rho_{av}$.  Upon using their Eq. (12) to calculate $\phi_s$ with the parameters of our simulations we find 
 $\phi_s \sim O(0.1)\%$, an order of magnitude or more smaller than our computed values. Another somewhat related paper is the study by~\citet{Sturtzetal21} which presents experimental results on the erosion of a particle-dense layer on the surface of a volumetrically heated liquid. In this paper, however, velocities were of the order of mm/s and the particle-liquid density contrast was of the order of $10^{-3}$. These parameters are so different from ours that no meaningful comparison seems possible. 

The gravitational potential energy of particles  suspended in Rayleigh-B\'{e}nard convection is ultimately dependent on the buoyant work done by gravity on the fluid and one might wonder why the efficiency estimated by Solomatov \& Stevenson is so much smaller than the values that we have calculated. In commenting the fact that, in the experiments by~\citet{MartinNokes89} on which they based in part their considerations, the suspended solid fraction was surprisingly small, the authors say ``it is more likely that ... [particle] reentrainment could not occur because the local mechanisms were not able to reentrain the particles, independently of our energetic criterion." The results that we have described suggest that the missing `local mechanisms' for re-suspension were dunes, which probably could not easily form in Martin \& Nokes's system due to the smallness of their particles relative to the boundary layer thickness on the cell bottom. 

Intense vortices localized above a sand bed are also known to re-suspend particles~\cite[see e.g.][]{Munroetal09}, and this is indeed at the root of the Shields mechanism of particle transport in the flow of air or water over sand~\cite[see e.g.][]{Ouriemietal07}. However, this mechanism relies on intense vorticity near the surface of the particle layer and would require a much larger Rayleigh number to be effective in Rayleigh-B\'{e}nard convection. At our $Ra$ vortices are not localized but span the entire cell height. 
% and, secondly, at $Ra =10^7$ it is not sufficiently intense. 
We have carried out some preliminary simulations with a single particle exposed to the same Rayleigh-B\'{e}nard flow and we have never observed a vortex-induced re-suspension process. 
%The only resuspension that takes place is due to the localized hot plume surrounding the particle as it gets heated up by the cell bottom.

%Our results show that the particle load that the fluid ultimately carries is not only limited by the available fluid energy, but also by the efficiency of the mechanism by which particles can be resuspended. 

\section{Conclusions}

The objective of this study was the elucidation, by computational means, of the mechanism by which particles subjected to  Rayleigh-B\'{e}nard convection become re-suspended in the flow once they have fallen to the bottom of the cell. This phenomenon cannot be studied by relying on the widely-used point-particle model for the obvious reason that particles with zero volume resting on the no-slip bottom are not subjected to any flow. The only option is to carry out particle-resolved simulations, which we have done at a Rayleigh number of $10^7$. (It may be noted in passing that the only other simulations of this nature currently available have been conducted for $Ra=10^5$, see e.g.~\cite{Takeuchietal19}.)

We have found that the flow near the bottom of the cell, directed from the touch-down spot of the descending plume toward the base of the ascending plume, entrains the particles resting on the bottom leading to the formation of heaps, or `dunes'  (figures~\ref{fqui} and~\ref{fsed150}). Other particles are entrained up the sides of the dunes thus acquiring a vertical velocity component which, once they reach the top of the dune, facilitates their entrainment by the ascending flow. This re-suspension mechanism, therefore, is not dependent on lift, but on drag forces. Since dunes are fairly stable on the time scale of the fluid circulation, they heat up (figure~\ref{fsed500}$b$) and strengthen the ascending plumes, which favors the re-suspension process. This mechanism was first described in a short 1993 paper by Solomatov {\em et al.} reporting the results of an experiment with polystyrene spheres in water but has only been mentioned occasionally in the literature~\cite[see e.g.][]{ElkinsTanton12}. 

The ready formation of dunes is favored when particles are not deeply embedded inside the convection momentum boundary layer. In our simulations the ratio of the particle diameter $d_p$ to the boundary-layer thickness $\delta$ 
(estimated using the wind-velocity, calculated as the average of the modulus of the vertical velocity on a plane at the cell mid-height) is $d_p/\delta \simeq 1.17$. %Use of the free-fall velocity gives instead $d_p/\delta \sim 1.4$. *
% is about 36\% of the particle diameter. 
The particles, therefore, are subjected to significant flow and the mechanism of dune formation of figure~\ref{fqui} can be quite effective. For much smaller values of $d_p/\delta$ re-suspension would be expected to be due to the Shields mechanism~\cite[see e.g.][]{Ouriemietal07} which requires a much larger Rayleigh number to become effective in Rayleigh-B\'{e}nard convection. 

%We may conclude that, although related, suspension and re-suspension are independent of each other and work in concert to determine the actual fraction of suspended particles. } 

In their study of particle suspension in natural convection, \citet{SolomatovStevenson93} focused on the energetics of particle suspension predicting a very small rate of conversion of fluid to particle gravitational energy. The results that we find for this ratio are significantly greater than this estimate. Our explanation for the difference is that, in addition to the energetic cost of keeping particles in suspension, a crucial role is played by the effectiveness of the process by which particles are re-suspended, which Solomatov \& Stevenson did not consider.

\begin{acknowledgments}
The authors acknowledge the use of the Carya and Sabine clusters and are grateful to Dr. J. Ebalunode of the UH Research Computing Data Core for his support.  %at the University of Houston to carry out this research.
\\
This study has been supported in part by the National Science Foundation Program CBET TTP under grant  No. 2053204.
\end{acknowledgments}

% \bigskip
% \noindent
% {\bf Declaration of Interests}. The authors report no conflict of interest. 

\setcounter{equation}{0}
\renewcommand{\theequation}{A.\arabic{equation}}

\appendix
\section{Collision modeling}

Particle-particle and particle-boundary collisions are handled by a soft-particle contact-force model, which provides the forces and the couples on the colliding particles, described in detail  in~\citet{SierakowskiProsperetti16}. The model  is based on work of~\citet{Tsujietal92} and~\citet{BarnockyDavis88}. The normal component of the collision force between two particles $\alpha$ and $\beta$ is given by
\be
{\bf F}_n = \left[-k_n(-h)^{3/2} -\eta_n ({\bf w}^\alpha -{\bf w}^\beta) \cdot{\bf n}\right]{\bf n}  \,,
\ee
in which $\bf n$ is a unit vector directed from the center of particle $\beta$ to that of particle $\alpha$, $(-h)$ is the amount of overlap between two geometric spheres separated by the distance between the particle centers, $k_n$ is given by 
\be
  \frac{\sqrt{2} }{3} \frac{E\sqrt{r_p}}{1-\sigma^2} \,,
  \ee
with $\sigma$ the Poisson ratio, $E$ Young's modulus, and $\eta_n$ a damping parameter defined by
\be
\eta_n =\alpha(e)\sqrt{\frac{1}{2}m_p k_n}\,(-h)^{1/4} \, .
\ee
Here $\alpha(e)$ is a function of the coefficient of restitution $e$ which can be approximated by~\cite{SierakowskiProsperetti16}
\be
 \alpha(e) = 2.22-2.26 \,e^{0.4} \, .
 \ee

For the tangential force we start from the velocity of the point of contact $\bf s$
\be
\frac{d{\bf s}}{dt} = ({\sf I}-{\bf nn})\cdot 
\left[ {\bf w}^\alpha -{\bf w}^\beta -\frac{1}{2}(2r_p+h)(\boldsymbol{\Omega}^\alpha-\boldsymbol{\Omega}^\beta)\times {\bf n} \right]\,,
\ee
with $\sf I$ the identity two-tensor and $\boldsymbol{\Omega}$ the particle angular velocity. Then we calculate the elastic tangential force
using
\be
{\bf F}_t =  \frac{d{\bf s}}{dt} k_t  \,,
\ee
 with the tangential stiffness $k_t$ given by Mindlin's theory~\cite[][]{Croweetal12} 
 \be
  k_t =  \frac{\sqrt{2}\,E}{(1-\sigma^2)(1+\sigma)} \sqrt{-r_p h}  \, .
  \ee
To ensure that the tangential force coincides with the tangential plane we project it back onto the current tangential plane by 
$({\sf I}-{\bf nn})\cdot {\bf F}_t \rightarrow {\bf F}_t$. Should ${\bf F}_t$ exceed that which can be supported by the frictional coefficient $\mu_f$,  $|{\bf F}_t| \leq \mu_F |{\bf F}_n|$, ${\bf F}_t$ is set to ${\bf F}_t=\mu_F |{\bf F}_n| ({\bf F}_t/|{\bf F}_t|)$.

\setcounter{equation}{0}
\renewcommand{\theequation}{B.\arabic{equation}}

\section{Particle-particle contact heat transfer }

Particles do not exchange heat only with the fluid but also with other particles when they collide.  This latter mode of energy transfer has been considered in~\citet{Takeuchietal19} and other studies from the same group in their simulations with $k_p/k_f=100$ and $Ra=10^5$. 
 
\citet{SunChen88} have studied the average heat transfer coefficient upon particle collisions in a gas fluidized bed. Since, in their situation, the effect of the interstitial fluid is essentially absent, one may expect that their estimate is the result of more energetic collisions than in a liquid, resulting in a stronger deformation of the particles and therefore an increased   heat transfer. On the other hand, the duration of the contact would be expected to be smaller, which would decrease the  heat transfer.  ~\citet{Davisetal86} considered the effect of the interstitial fluid and came up with the same scaling (\ref{contt}) as Sun \& Chen up to constants of order one which will be seen to be unimportant for the matter of present concern.

For equal particles, Sun \& Chen's Eq. (24) reduces to
 \be
  h_c =  0.435 C \frac{N_p v}{V}  \rho_p c_{pp} A_c\sqrt{D_{p,th}t_c} \,,
  \ee
  in which $C>1$ is a coefficient of order 1, $N_p$ the total number of particles in the domain of volume $V$, $v$ the typical collision velocity, $A_c$ the area of the contact disk and $t_c$ the duration of the contact. For equal particles the area is given by 
  \be
   A_c = \pi \left(\frac{20\pi \rho_p r_p^5(1-\sigma^2)}{64E}\right)^{2/5} v^{4/5} \, ,
   \ee
  with $\sigma$ the Poisson ratio and $E$ Young's modulus, and the duration of the contact by
  \be
   t_c= 3.38 \left(\frac{20 \pi\rho_p r_p^3 (1-\sigma^2)}{16 E}\right)^{2/5}  (r_p v)^{-1/5} \, .
   \label{contt}
   \ee

We take the Stokes terminal velocity $U_{St}= (\rho_p/\rho_f-1)gd_p^2/(18 \nu_f)$ as the characteristic collision velocity $v$, with the corresponding Reynolds number $Re_{St}=d_pU_{st}/\nu_f$ and find (\ref{collt}) for $\nu_f t_c/d_p^2$ and 
\be
 \frac{A_c}{\pi r_p^2}=  
 \left( \frac{5\pi}{5184}  \frac{\rho_p}{\rho_f} \left(\frac{\rho_p}{\rho_f}-1\right)^2  \frac{1}{E_*} \right)^{2/5}\,.
 \ee
% \be
  %\frac{\nu_f t_c}{d_p^2} = 1.69  \left(\frac{5\pi}{1296}\right)^{2/5}  \frac{1 }{Re_{St}} 
 %\left(\frac{\rho_p}{\rho_f} \left(\frac{\rho_p}{\rho_f}-1\right)^2  \frac{1}{E_*} \right)^{2/5} \,.
 %\ee
 The Nusselt number for collision heat transfer can then be written as
 \bea
   Nu_c &=&\frac{d_p h_c}{k_f} = 0.738 \left(\frac{5\pi}{2592}\right)^{3/5} \frac{C \alpha}{E_*^{3/5}}
 \left(\frac{\rho_p}{\rho_f}\right)^{11/10}  \left(\frac{\rho_p}{\rho_f}-1\right)^{6/5} 
   \sqrt{Pr \, Re_{St}\,\frac{k_p}{k_f} \frac{c_{pp}}{c_{pf}} } \,. 
   \nonumber
 \eea

With the parameters of the simulations, $C=1$ and $k_p/k_f=1$ this relation gives $Nu_c \sim 0.0046$, about three orders of magnitude smaller than the value $Nu_p\sim 3.9$ of the Nusselt number for fluid-particle heat exchange of the present simulations (see Table~\ref{tabprop}). In order to achieve a comparable Nusselt number it would be necessary that $k_p/k_f\sim 10^6$, very much larger than the particle conductivity necessary to justify the lumped-capacitance approximation (\ref{parten}).

\bibliography{refs}

\end{document}